\documentclass[onecolumn]{aa}
\usepackage{epsfig}
\input psfig.sty
\begin{document}

\title{Cosmic distance duality relation and the shape of galaxy clusters}
\author{
R.\ F.\ L.\ Holanda\inst{1}\thanks{\email{holanda@astro.iag.usp.br}}
\and
J.\ A.\ S.\ Lima\inst{1}\thanks{\email{limajas@astro.iag.usp.br}}
\and
M.\ B.\ Ribeiro\inst{2}\thanks{\email{mbr@if.ufrj.br}}
      }

\institute{
Departamento de Astronomia, Instituto Astron\^{o}mico e Geof\'{\i}sico,
Universidade de S\~{a}o Paulo -- USP, S\~{a}o Paulo, Brazil
\and
Instituto de F\'{\i}sica, Universidade Federal do Rio de Janeiro -- UFRJ,
Rio de Janeiro, Brazil
         }


\date{}

\abstract {Observations in the cosmological domain are heavily dependent on the
validity of the cosmic distance-duality (DD) relation, $D_L(z)
(1 + z)^{2}/D_{A}(z) = 1$, an exact result required by the Etherington
reciprocity theorem where $D_L(z)$ and $D_A(z)$ are, respectively, the luminosity and
angular diameter distances. In the limit of very small redshifts $D_A(z)
= D_L(z)$ and this ratio is trivially satisfied. 
Measurements of Sunyaev-Zeldovich effect (SZE) and X-rays combined with
the DD relation have been used to determine $D_A(z)$ from galaxy clusters. This
combination offers the possibility of testing the validity of the DD
relation, as well as determining which physical processes occur in galaxy clusters
via their shapes.} {  { We use WMAP (7 years) results  by fixing the conventional
$\Lambda$CDM model} to verify the consistence between the
validity of DD relation  and different assumptions about galaxy
cluster geometries usually adopted in the literature.} {We assume that $\eta$ is a function of the
redshift parametrized by two different relations: $\eta(z) = 1
+ \eta_{0}z$, and $\eta(z)=1 + \eta_{0}z/(1+z)$, where $\eta_0$ is a
constant parameter quantifying the possible departure from the
strict validity of the DD relation. In order to determine the
probability density function (PDF) of $\eta_{0}$, we consider the
angular diameter distances from galaxy clusters recently studied by
two different groups by assuming elliptical (isothermal) and
spherical (non-isothermal) $\beta$ models. The strict
validity of the DD relation will occur only if the maximum value of
$\eta_{0}$ PDF is centered on $\eta_{0}=0$.} {It was found that the
elliptical $\beta$ model is in good agreement with the data, showing
no violation of the DD relation (PDF peaked close to
$\eta_0=0$ at $1\sigma$), while the spherical (non-isothermal) one
is only marginally compatible at $3\sigma$.}{The present results
derived by combining the SZE and X-ray surface brightness data from
galaxy clusters with the latest WMAP results (7-years) favors the
elliptical geometry for galaxy clusters. It is remarkable
that a local property like the geometry of galaxy clusters might
be constrained by a global argument provided by the
cosmic DD relation.}

\authorrunning
\titlerunning
\maketitle

\section{Introduction}

The most  useful distances in cosmology are the luminosity distance, $D_L(z)$, and the angular-diameter distance, $D_A(z)$. 
The expressions of both distances depend on the world models, but the relationship between them, namely
  \begin{equation}
 \frac{D_{\scriptstyle L}}{D_{\scriptstyle A}}{(1+z)}^{-2}=1
 \label{rec0}
\end{equation}
is valid for arbitrary spacetimes, a result usually  referred to as distance-duality (DD) relation.   

The above expression can easily be deduced in the context of
Friedmann-Robertson-Walker (FRW) cosmologies (Weinberg
1972).  However, as originally proven by Etherington (1933), it 
depends neither on Einstein field equations nor the nature of matter content filling
the spacetime. The proof depends crucially on photon conservation 
(transparency of the cosmic medium) and that sources and observers are linked by null geodesics in a Riemannian spacetime.

The DD relation plays an essential role ranging from gravitational
lensing studies to analyses of the cosmic microwave blackbody
radiation (CMBR) observations, as well as for galaxy and galaxy
cluster observations (Schneidder, Ehlers \& Falco 1999; Komatsu et
al.\ 2011; Lima, Cunha \& Alcaniz 2003; Cunha, Marassi \& Lima 2007;
Ribeiro 1992, 2005; Ribeiro \& Stoeger 2003). Indeed, any 
observational deviation from Eq. (\ref{rec0}) would be a theoretical
catastrophe thereby igniting a major crises in observational cosmology (Ellis 1971, 2007).  

Although  taken for granted in virtually all analyses in cosmology,
the DD relation is in principle testable by means of astronomical
observations. One may  assume a redshift dependence of the form 
\begin{equation}
 \frac{D_{\scriptstyle L}}{D_{\scriptstyle A}}{(1+z)}^{-2}= \eta(z),
 \label{rec}
\end{equation}
where $\eta(z)$ quantifies a possible epoch-dependent departure
from the standard photon conserving scenario ($\eta=1$). 

Basset \& Kuns (2004) used both supernovae Ia
data as measurements of the luminosity distance $D_{\scriptstyle L}$
and the estimated $D_{\scriptstyle A}$ of FRIIb radio galaxies
(Daly \& Djorgovski 2003) and ultra compact radio sources (Gurvitz
1994, 1999; Lima \& Alcaniz 2000, 2002; Santos \& Lima 2008) in adopting 
this kind of approach to test possible new physics. Any source 
of attenuation (``gray"' intergalactic dust) 
or exotic photon interaction must violate the DD relation 
(More et al. 2009, Avgoustidis et al. 2010),  
 thereby providing new consistency checks of cosmological models. 

On the other hand,  observations of the Sunyaev-Zeldovich
effect (SZE) from galaxy clusters are becoming a powerful tool in
cosmology (Sunyaev \& Zeldovich 1972; Cavaliere \& Fusco-Fermiano
1978, De Filippis et al. 2005, Cunha et al. 2007; Nord et al. 2009;
Basu et al. 2010). The
combination of SZE and X-ray provides the $D_A(z)$ of galaxy
clusters, hence can be used to constrain some cosmological
parameters. However, Uzan, Aghanim \& Mellier (2004) argued that this
technique  is strongly dependent on the validity of the DD relation.
When the DD relation does not hold ($\eta \neq 1$), the
observationally determined angular distance must be replaced by the value (in the Uzan et al.\ (2004) 
notation the correcting term is $\eta^{-2}$)
\begin{equation}
D^{\: data}_{A}(z)=D_{A}(z)\eta^{2},
\label{4}
\end{equation}
this quantity reduces to the conventional angular diameter distance (assuming cosmic transparency) only when the DD
relation is strictly valid ($\eta=1$). To quantify the
$\eta$ parameter, Uzan et al.\ (2004) obtained $D_A(z)$ as given by the 
cosmic concordance model (Spergel et al. 2003), whereas for $D^{\:
data}_{A}(z)$ they considered 18  angular diameters from the
Reese {\it et al.\ } (2002) galaxy cluster sample for which a spherically symmetric
cluster geometry was assumed. By assuming $\eta$ to be constant, their
statistical analysis provided $\eta = 0.91^{+ 0.04}_{-0.04}$
(1$\sigma$) and is therefore only marginally consistent with the
standard result, $\eta=1$.


De Bernardis, Giusarma \& Melchiorri (2006) also searched
for deviations from the DD relation by using the {\bf $D^{\: data}_{A}(z)$} from
galaxy clusters provided by the sample of Bonamente {\it et al.\ }
(2006). They found  a non violation of the DD relation in the
framework of the cosmic concordance $\Lambda$CDM model. {
Avgoustidis {et al.\ }(2010) adopted a DD relation of the
form $D_L=D_A(1+z)^{2+\epsilon}$ to constrain the cosmic opacity
by combining the  SN Type Ia data compilation of Kowalski { et al.}
(2008) with the latest measurements of the Hubble expansion at
redshifts in the range $0 < z < 2$ (Stern { et al.} 2009). By
working in the context of a flat $\Lambda$CDM model, they found
$\epsilon=-0.04_{-0.07}^{+0.08}$ (2$\sigma$).}

In the past few years, many studies based on 
{\it Chandra } and {\it XMM} observations have shown that 
in general galaxy clusters exhibit
elliptical surface brightness maps. Simulations have also predicted
that dark matter halos show axis ratios typically of the order of
$\approx 0.8$ (Wang \& White 2009), thereby  disproving the spherical
geometry assumption usually adopted (Reiprich \& Boringer
2002; Bonamente et al.\ 2006, Shang, Haiman \& Verdi 2009). In this line, the first determination of the
intrinsic three-dimensional (3D) shapes of galaxy clusters was
presented by Morandi, Pedersen \& Limousin (2010) by combining
X-ray, weak-lensing, and strong-lensing observations. They studied
the galaxy cluster MACS J1423.8+2404 and found a
tri-axial galaxy cluster geometry with DM halo axial ratios $1.53
\pm 0.15$ and $1.44 \pm 0.07$ on the plane of the sky and along the
line of sight, respectively.

In this letter, we take the validity of the DD relation for
granted  to access the galaxy cluster morphology. The 
values of $D_A(z)$ are obtained from the WMAP (7 years) results by
fixing the conventional flat $\Lambda$CDM model whereas the
observational measurements $D_A^{\: data}(z)$ are the angular
diameter distances from galaxy clusters obtained via SZE plus X-ray
techniques. These samples differ in terms of the assumptions concerning the
possible cluster geometries of elliptical and spherical models. Our analysis is based on two parametric
representations of $\eta(z)$  defined by Eq. \ref{rec} (or Eq.
\ref{4}), namely
\\
\\
\hspace{1.0cm} I. $\eta (z) = 1 + \eta_{0} z$ \, \, and \, \, II.
$\eta (z) = 1 + \eta_{0}z/(1+z)$.
\\
\\
\noindent  The first expression is a continuous and smooth
one-parameter linear expansion, whereas the second one includes a
possible epoch-dependent correction, which avoids the divergence at
extremely high z. These 
deformations of the DD 
relation effectively parametrize our ignorance of the underlying process 
responsible for its possible violation. 
However, we emphasize that these
expressions are very simple and have several advantages such as 
a manageable one-dimensional phase space and a good
sensitivity to observational data. Clearly, the second
parametrization can also be rewritten as $\eta (z) = 1 +
\eta_{0}(1-a)$, where $a(z)= (1 + z)^{-1}$ is the cosmic scale
factor. This represents an improvement with respect to the linear
parametrization, since the DD relation becomes bounded regardless of
the redshift values. It will become more useful once
higher redshift clusters data become available.

The above parametrizations are clearly inspired by similar
expressions for the $\omega(z)$-equation of state parameter of dark energy models (Padmanabhan \&
Choudury 2003; Linder 2003; Cunha, Marassi \& Santos 2007; Silva,
Alcaniz \& Lima 2007).  In the limit of extremely low redshifts
($z<<1$), we have $\eta = 1$ and $D_{L} = D_{A}$ as should be
expected, and, more important for our subsequent analysis, the value
$\eta_0=0$ must be favored by the Etherington result. In other
words, for a given data set, the likelihood of $\eta_0$ must 
peak at $\eta_0=0$ to satisfy the cosmic
relation. As we shall see, for those accepting the strict
validity of the standard DD relation, our analysis suggests that
galaxy clusters have an elliptical geometry. In principle, this
kind of result is an interesting example of how a cosmological
(global) condition correlates with the local physics.
\begin{figure}
   \centering
       \includegraphics[width=0.55\linewidth]{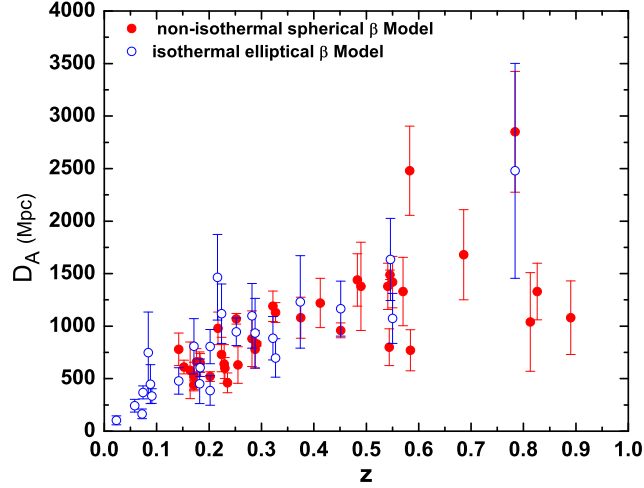}
   \caption{Galaxy clusters data. The open (blue) and filled (red)
            circles with the associated error bars represent, respectively,
          the De Filippis et al.\ (2005) and Bonamente et al.\
         (2006) samples.}
   \label{fig1}
\end{figure}

\section{Galaxy cluster samples }\label{sec:Sample}

To constrain the values of $\eta_{0}$, we consider
two galaxy cluster samples for which {\bf $D_A(z)^{data}$} were derived by
combining their SZE and X-ray surface brightness observations.

The first sample is formed by 25 galaxy clusters compiled by De
Filippis {{et al.}} (2005).  By using an isothermal elliptical
$\beta$ model to describe the clusters, the 
{\bf $D_A(z)^{data}$} was derived for two subsamples discussed in the literature. 
The first one, compiled by Reese et al.\ (2002), is a
selection of 18 galaxy clusters distributed over the redshift
interval $0.14 < z < 0.8$. The second subsample of Mason et
al.\ (2001) has 7 clusters from the X-ray limited flux sample of
Ebeling et al.\ (1996). These 25 pieces of data are referred to as the elliptical sample.

The second sample is defined by the 38 galaxy clusters observed by
Bonamente {et al.} (2006), where the cluster plasma and dark matter
distributions were analyzed assuming hydrostatic equilibrium model
and  spherical symmetry. This sample
consists of clusters that have both X-ray data from the {\it Chandra
Observatory} and SZE data from the BIMA/OVRO SZE imaging project,
which uses the Berkeley-Illinois-Maryland Association (BIMA) and
Owens Valley radio observatory (OVRO) interferometers to image the
SZE. This dataset is termed  the spherical sample.

In Fig.\ \ref{fig1}, we plot the elliptical and spherical galaxy
cluster samples. Some authors have adopted these samples to estimate
the galaxy cluster distances and measure the Hubble parameter
by means of the SZE/X-ray technique (Bonamente et al. 2006; Cunha,
Marassi \& Lima 2007). However, since these samples are endowed with
different geometric assumptions, our main interest here is to
confront these underlying hypotheses  with the validity of the DD
relation.
 
At present, there is no convincing evidence for deviations from the minimal
cosmic concordance flat model (Komatsu et al.\
2011, Percival et al.\ 2010, Riess et al.\ 2009).  In this case,
the angular diameter distance  is given by (Lima et al.
2003; Cunha, Marassi \& Lima 2007)
\begin{equation}
{D}_{A}(z;h,\Omega_m) = \frac{3000h^{-1}}{(1 +
z)}\int_{o}^{z}\frac{dz'}{{\cal{H}}(z';\Omega_m)}\,\mbox{Mpc},
\label{eq1}
\end{equation}
where $h=H_0/100$ km s$^{-1}$ Mpc$^{-1}$ and the dimensionless
function ${\cal{H}}(z';\Omega_m)$ is given by
\begin{equation}
{\cal{H}} = \left[\Omega_m(1 + z')^{3} + (1 - \Omega_{m})\right]^{1/2}.
\label{eq2}
\end{equation}

\begin{figure}
\centering
{\includegraphics[width=90mm, angle=0]{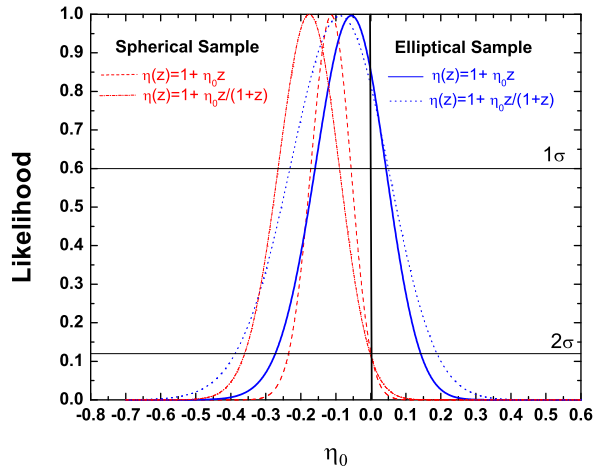}

\hskip 0.1in} \caption{The likelihood distribution functions for
spherical (dotted-dashed and dashed red lines) and elliptical (dotted and solid blue lines) cases.  
See the text for values of $\eta_0$.}
\label{fig:Analysis}
\end{figure}

In our statistical analysis (see next section), the 
$D_{A}(z)$ value was calculated by using the cosmological parameters
given in the joint analysis  carried out by Komatsu et al.\
(2011), namely $\Omega_{\Lambda}=(1 - \Omega_{m})=0.728\pm0.015$ and
$h=0.704\pm0.013$.

\section{Analysis and Results}\label{sec:analysis }

We estimate the $\eta_{0}$ parameter for each sample and both parameterizations of $\eta(z)$,
namely, $\eta(z) = 1+\eta_{0}z$ and $\eta(z) = 1+\eta_{0}z/(1+z)$. To begin
with,  we evaluate the likelihood distribution function
$e^{-\chi^{2}/2}$ , where
\begin{equation}
\label{chi2} \chi^{2} = \sum_{z}\frac{{\left[\eta^{2}(z) -
\eta^{2}_{obs}(z) \right] }^{2}}{\sigma_{\eta^{2}_{obs}} }
\end{equation}
and $\eta^{2}_{obs}(z) = D^{\: data}_{A}(z)/D_{A}(z)$ (see Eq. (\ref{4})). The
statistical and systematic errors have been discussed by many authors 
(Mason et al, 2001; Reese et al. 2002, 2004; Bonamente et al. 2006). 
Statistical error contributions for galaxy clusters are: SZE
point sources $\pm 8$\%, X-ray background $\pm 2$\%, Galactic N$_{H}$
$\leq \pm 1\%$, $\pm 15$\% for cluster asphericity, $\pm 8$\%
kinetic SZ and for CMBR anisotropy $\leq \pm 2\%$. On the other hand,
the estimates of systematic effects are: SZ calibration
$\pm 8$\%, X-ray flux calibration $\pm 5$\%, radio halos $+3$\%, and
X-ray temperature calibration $\pm 7.5$\%. One
may show that typical statistical errors can reach nearly $20$\%,
in agreement with other works (Mason et al. 2001; Reese et al. 2002,
2004), whereas for systematics we also find typical errors around
+12.4\% and $-$12\%  (see also table 3 in Bonamente et al. 2006). 

In Fig.\ \ref{fig:Analysis}, we plot the likelihood distribution
function for each galaxy cluster data. The basic results are:
\begin{itemize}
\item For the elliptical  sample, we obtain $\eta_{0} =
     -0.056^{+ 0.1}_{-0.1}$ ($\chi_{d.o.f.}^2 = 0.98$) and $\eta_{0}
     = -0.088^{+ 0.14}_{-0.14}$ ($\chi_{d.o.f.}^2 = 0.97$) in
     $68\%$ c.l.\ for a linear (blue solid line) and  non linear
     (blue dotted line) parametrization, respectively.
\item For the spherical sample, we obtain $\eta_{0} = -0.12^{+
     0.055}_{-0.055}$ ($\chi_{d.o.f.}^2 = 0.85$) and $\eta_{0} =
     -0.175^{+0.083}_{- 0.083}$ ($\chi_{d.o.f.}^2 = 0.84$) in
     $68\%$ c.l.\ for a linear (red dashed line) and  non linear
     (red dotted dashed line) parametrization.
\end{itemize}
The uncertainties in $ \eta_0$ include (in quadrature) the systematic plus
statistical errors. We can see that the $D^{\:data}_A(z)$
inferred  based on the elliptical description agrees
with the DD relation, whereas the other case, in which a spherical
$\beta$ model was assumed to describe the clusters, is only
marginally compatible with the DD relation. This result remains
valid even when only clusters with $z>0.1$ are considered in the
elliptical sample. In this situation, we obtain $\eta_{0} =
-0.044^{+ 0.1}_{- 0.1}$ ($\chi_{d.o.f.}^2 = 0.94$) for the linear
parametrization, and $\eta_{0} = -0.07^{+ 0.14}_{- 0.14}$
($\chi_{d.o.f.}^2 = 0.93$) within 1$\sigma$ in the non-linear case.

It thus follows that we have found no evidence of violation of the DD
relation when the  elliptical case is considered. However, the
same kind of analysis shown in Fig. 2 points to a contradiction with
the spherical symmetry hypothesis assumed in the Bonamente et al.\
(2006) sample.

\section{Conclusions}\label{sec:Conclusions}

We have explored some consequences of a deformed distance
duality relation, $\eta(z) = D_{L}(1+z)^{-2}/D_{A}$, based on
observations of Sunyaev-Zeldovich effect and X-ray from galaxy
clusters. The consistency between the
strict validity of the standard relation ($\eta(z) \equiv 1$) and the
assumptions regarding the geometry used to describe the galaxy
clusters (elliptical and spherical $\beta$ models) has been discussed. The $\eta(z)$
function was parametrized in two distinct forms, $\eta = 1 +
\eta_{0}z$ and $\eta = 1 + \eta_{0}z/(1+z)$, where $\eta_0$ is
a constant parameter quantifying  a possible departure from the
strict validity of the duality relation. The basic idea pursued in
this work is a simple one. The likelihood of the free parameter
appearing in the proposed expressions for $\eta(z)$ should peak
around $\eta_0 =0$ when the distance duality relation is strictly
obeyed.

By comparing the De Filippis et al.\ (2005) (elliptical isothermal
$\beta$ model) and Bonamente  et al.\ (2006) (spherical
non-isothermal $\beta$ model) samples with  $D_{A}(z)$ obtained from 
$\Lambda$CDM (WMAP7), we show that the
elliptical geometry is more consistent with no violation of the
duality relation.
The uncertainties in $ \eta_0$ included the
systematic plus statistical errors from cluster data. In the case
of an elliptical sample (see Fig.\ \ref{fig:Analysis}), we found that
$\eta_{0} = -0.056^{+0.1}_{- 0.1}$ and $\eta_{0} = -0.088^{+
0.14}_{- 0.14}$ for the linear and non-linear parametrization,
respectively. However, the spherical sample (see
Fig. 2) is only marginally compatible with $\eta_{0} =
-0.12^{+0.055}_{-0.055}$ and $\eta_{0} = -0.175^{+ 0.083}_{-0.083}$
for linear and non-linear parametrization, respectively. Our
analysis reveals that the elliptical model is compatible with the
duality relation validity at 1$\sigma$, whereas the spherical model
is only marginally compatible at 3$\sigma$.

At this point, it is interesting to compare our results
with those obtained by following a complementary
approach (Holanda, Lima \& Ribeiro 2010). The $\eta(z)$ function
there was also parametrized as in the present work. However, the
overall discussion was based on a model-independent cosmological
test  by considering $D_A(z)$ from galaxy clusters and the
luminosity distances given by two sub-samples of SNe Ia taken from
the constitution data (Hicken et al. 2009). Both analyse are consistent
with each other and suggest that the elliptical model is more
compatible with the validity of the standard duality relation than
the spherical case.

 Summarizing, the statistical analysis presented here provides additional
evidence that the true geometry of clusters has an elliptical form.
In principle, it is remarkable that a local property (the geometry
of galaxy clusters) might be constrained by a global argument such as
the one provided by the cosmological distance duality relation. In the near
future, as more and larger data sets with smaller statistical and
systematic uncertainties become available, the method proposed here
(based on the validity of the distance duality relation) can improve
the limits on the measurements of cluster geometries.

\centerline{\bf Acknowledgments}

The authors are grateful to an anonymous referee for helpful comments
and suggestions that improved the original version of the work.  We  also thank Antonio Guimar\~aes for helpful
discussions.  RFLH is supported by FAPESP (No. 07/52912-2), JASL is
partially supported by CNPq (No. 304792/2003-9) and FAPESP (No.
04/13668-0) and MBR is partially supported by FAPERJ.

\label{lastpage}

\begin{thebibliography}{99}
\bibitem{lverde}{ Avgoustidis, A., Burrage, C., Redondo, J., Verde,
        L., \& Jimenez, R. 2010, JCAP, 1010, 024, [arXiv:1004.2053]}
\bibitem{bk04} Basset, B.A., \& Kunz, M. 2004, PRD, 69,
        101305
\bibitem{basu10} { Basu, K., et al. 2010, A\&A, 519, A29 [arXiv:0911.3905v3]}
\bibitem{Boname06} Bonamente, M., et al. 2006, ApJ, 647, 25
\bibitem{caval} Cavaliere, A., \& Fusco-Fermiano, R. 1978, A\&A., 667, 70
\bibitem{CMS07} Cunha, J. V., Marassi, L., Santos, R.C. 2007, IJMPD, 16,
        403 
\bibitem{CML07}  Cunha, J.\ V., Marassi, L., Lima, J.\ A.\ S. 2007,
        MNRAS, 379, L1 [astro-ph/0611934]
\bibitem{daly} Daly, R.\ A., \& Djorgovski, S. G. 2003, ApJ, 597, 9
\bibitem{bem06} De Bernardis, F., Giusarma, E., \& Melchiorri A. 2006,
        IJMPD, 15, 759 [arXiv:gr-qc/0606029v1]
\bibitem{DeFilippis05} De Filippis, E., Sereno, M., Bautz, M.\ W., \&
        Longo G.\ 2005, ApJ, 625, 108
\bibitem{ebelin} Ebeling, H., et al.\ 1996, MNRAS, 281, 799
\bibitem{ellis71} Ellis, G. F. R. 1971, ``Relativistic Cosmology'',
        Proc. Int. School Phys. Enrico Fermi, R. K. Sachs (ed.),
        pp.\ 104-182 (Academic Press: New York) reprinted in GRG
     2009, 41, 581
\bibitem{ellis07} Ellis, G.\ F.\ R.\ 2007, GRG, 39, 1047
\bibitem{eth33} Etherington, I.\ M.\ H.\ 1933, Phil.\ Mag., 15, 761;
        reprinted in 2007, GRG, 39, 1055
\bibitem{G94} Gurvitz, L.\ I.\ 1994, ApJ, 425, 442 [arXiv:1005.4458]
\bibitem{g99} Gurvitz, L. I., Kellermann, K. I., \& Frey, S. 1999, A\&A, 342, 378
\bibitem{Hicken}Hicken, M., et al. 2009, ApJ, 700, 1097
\bibitem{holapjl}{ Holanda, R.\ F.\ L., Lima, J.\ A.\ S.\ \& Ribeiro,
        M.\ B.\ 2010, ApJL, 722, L233 [arXiv:1005.4458]}
\bibitem{union}{ Kowalski, M., et al.\ 2008, ApJ, 749, 686}
\bibitem{komatsu} Komatsu, E., et al.\ 2011, ApJS, 192, 18 [arXiv:1001.4538]
        (WMAP collaboration) 

\bibitem{LA00} Lima, J.\ A. S., \& Alcaniz, J. S. 2000, A\&A., 357,
        393 [astro-ph/0003189]
\bibitem{LA02} Lima, J. A.\ S., \& Alcaniz, J. S. 2002, ApJ, 566, 15
        [astro-ph/0109047]
\bibitem{LAC03} Lima, J.\ A.\ S., Cunha, J. V., \& Alcaniz, J.\ S.\
        2003, Phys.\ Rev.\ D, 68, 023510 [astro-ph/0303388]
\bibitem{linder04} Linder, E.\ V.\ 2003, PRL, 90, 091301
\bibitem{more09} {More, S., Bovy, J., Hogg, D. W. 2009, ApJ, 696, 1727}
\bibitem{Mas01}  Mason, B.\ S., et al. 2001, ApJ, 555, L11
\bibitem{morandi}{ Morandi, A., Pedersen, K., \& Limousin, M.\ 2010,
        ApJ, 713, 491}
\bibitem{Nord09} { Nord, M., et al. 2009, A\&A, 506, 623 [arXiv:0902.2131v2]}
\bibitem{pad04} Padmanabhan, T., \& Choudury R. 2003, MNRAS, 344, 823
\bibitem{percival} Percival, W., et al. 2010, MNRAS, 401, 2148
\bibitem{rb02}{ Reiprich, T. H., \& Bohringer, H. 2002, ApJL, 567, 716}
\bibitem{rib92}{ Ribeiro, M.\ B. 1992, ApJ, 388, 1 [arXiv:0807.0866]}
\bibitem{rib05}{ Ribeiro, M.\ B. 2005, A\&A., 429, 65
        [astro-ph/0408316]}
\bibitem{RS03} Ribeiro, M.\ B., \& Stoeger, W.\ R.\ 2003, ApJ, 592, 1
        [astro-ph/0304094]
\bibitem{Reese02}  Reese, E.\ D., et al. 2002, ApJ, 581, 53
\bibitem{Reese04}  Reese, E.\ D.\ 2004, in Measuring and Modeling the
        Universe, ed. W.\ L.\ Freedman (CUP) p. 138 [astro-ph/0306073]
\bibitem{riess} Riess, A., et al. 2009, ApJ, 116, 1009
\bibitem{SL08} Santos, R.\ C., Lima, J.\ A.\ S.\ 2008, PRD, 77, 083505
        [arXiv:0803.1865]
\bibitem{SEF99} Schneider, P.,  Ehlers, J.\ \& Falco, E.\ E.,
        Gravitational Lenses (Springer-Verlag, Berlin, 1992)
\bibitem{SHL09} { Shang, C., Haiman, Z., \& Verde, L. 2009, MNRAS, 400, 2, 1085 [arXiV:0908.2012v1]}
\bibitem{SAL07} Silva, R., Alcaniz, J.\ S., \& Lima, J.\ A.\ S.\ 2007,
        IJMPD, 16, 469
\bibitem{spergel03} Spergel, D.\ N., et al. 2003, ApJS, 148, 175
\bibitem{lverde2} Stern, D., Jimenez, R., Verde, L., Kamionkowski, M.\
        \& Stanford, S.\ A.\ [astro-ph/09073149]
\bibitem{SunZel72} Sunyaev, R.\ A., \& Zel'dovich Ya.B.\ 1972,
        Comments Astrophys.\ Space Phys., 4, 173
\bibitem{uzan} Uzan, J.\ P., \&  Aghanim, N., Mellier Y. 2004, PRD
        70, 083533
\bibitem{wang}{ Wang, J., \& White, S.\ D.\ M.\ 2009, MNRAS, 396,
        709}
\bibitem{wein} Weinberg, S., Gravitation and Cosmology (Wiley, New
        York, 1972)
\end{thebibliography}
\end{document}